\documentclass[nofootinbib,showpacs,showkeys,prd,superscriptaddress]{revtex4}

\usepackage{amsmath}
\usepackage{amsfonts}
\usepackage{amssymb}
\usepackage{graphicx}

\begin{document}

\title{Bounding $f(R)$ gravity by  particle production rate}

\author{Salvatore Capozziello}
\email{ capozziello@na.infn.it}
\affiliation{Dipartimento di Fisica, Universit\`a di Napoli ''Federico II'', Via Cinthia, Napoli, Italy.}
\affiliation{Istituto Nazionale di Fisica Nucleare (INFN), Sez. di Napoli, Via Cinthia, Napoli, Italy.}
\affiliation{Gran Sasso Science Institute (GSSI), Viale F. Crispi, 7, I-67100, L'Aquila, Italy.}
\affiliation{Tomsk State Pedagogical University, ul. Kievskaya, 60, 634061 Tomsk, Russia.}

\author{Orlando Luongo}
\email{luongo@na.infn.it}
\affiliation{Dipartimento di Fisica, Universit\`a di Napoli ''Federico II'', Via Cinthia, Napoli, Italy.}
\affiliation{Istituto Nazionale di Fisica Nucleare (INFN), Sez. di Napoli, Via Cinthia, Napoli, Italy.}
\affiliation{Department of Mathematics and Applied Mathematics, University of Cape Town, Rondebosch 7701, Cape Town, South Africa.}
\affiliation{Astrophysics, Cosmology and Gravity Centre (ACGC), University of Cape Town, Rondebosch 7701, Cape Town, South Africa.}

\author{Mariacristina Paolella}
\email{paolella@na.infn.it}
\affiliation{Dipartimento di Fisica, Universit\`a di Napoli ''Federico II'', Via Cinthia, Napoli, Italy.}
\affiliation{Istituto Nazionale di Fisica Nucleare (INFN), Sez. di Napoli, Via Cinthia, Napoli, Italy.}

\begin{abstract}
Several models of  $f(R)$ gravity have been proposed in order to address the dark side problem in cosmology. However,  these models should be constrained also at ultraviolet scales  in order to achieve some correct fundamental interpretation. Here we analyze this  possibility comparing  quantum vacuum states  in given $f(R)$ cosmological backgrounds. Specifically,  we compare the Bogolubov transformations associated to different vacuum states  for some
   $f(R)$ models.  The procedure consists in  fixing  the $f(R)$ free parameters by requiring that the Bogolubov coefficients can be correspondingly minimized to be in agreement with both high redshift observations and quantum field theory predictions.  In such a way, the  particle production is related to the value of  the Hubble parameter and then to the given $f(R)$ model.  The approach is developed in both metric and Palatini formalism.
\end{abstract}

\pacs{04.50.Kd; 98.80.Qc; 95.36.+x; 98.80.-k}
\keywords{Alternative Theories of Gravity; Cosmology; Quantum Field Theory; Dark Energy.}

\date{\today}

\maketitle


\section{Introduction}

Extensions of General Relativity can contribute to achieve  a comprehensive cosmological picture by giving a geometrical interpretation of the dark side \cite{numero1}. Although General Relativity is accurately bounded at the Solar System scales \cite{zero1,zero2}, several cosmological indications seem to point out the need of extending the  Hilbert-Einstein action in order to achieve a comprehensive and self-consistent description of the universe expansion history, i.e. at far infrared scales \cite{miareview,alterr,alterr1}. Rephrasing it differently, we wonder whether General Relativity is effectively the final paradigm to address consistently the  \emph{universe dynamical problem} \cite{uno3,due,due1}. In particular, the issue related to the existence of a dark energy, which drives the late-time dynamics, needs the introduction of additional fluids capable of dominating over the standard  pressureless  matter \cite{galaxy}. Moreover, the issue of dark matter is needed  to address the   clustering structures and  asks for  some additional  particles  resulting hard to find out at fundamental level by direct and indirect detection  \cite{tre, tre1,tre2,tre3,lucia}.

 Among the possible proposals,  $f(R)$ gravity seems quite promising as a straightforward extension of General Relativity since the strict request of linearity in the Ricci scalar $R$ of the Hilbert-Einstein action is relaxed. The paradigm consists in the fact  that observations and phenomenology, in principle, could fix the  action of gravitational  interaction whose action is assumed as  a generic function of the curvature invariants.  In this perspective, the simplest generalization is \cite{ciris}
 \begin{equation}\label{eq: action}
{\cal{A}} =\frac{1}{2} \int{d^4x \sqrt{-g} \left [ f(R) + {\cal{L}}_{m} \right]}\,,
\end{equation}
where $g$ is the determinant of the metric $g_{\mu\nu}$ and ${\cal L}_{m}$ is the standard perfect fluid matter Lagrangian.
 We adopt the conventions $8 \pi G = c = 1$.
The
variation of (\ref{eq: action}) with  respect to  $g_{\mu\nu}$ gives
the field equations \cite{OdintsovPR,6,uno3,10,libroSV,libroSF}:

\begin{equation}
f'(R)R_{\mu\nu}-\frac{1}{2}f(R) g_{\mu\nu}-\left[\nabla_{\mu} \nabla_{\nu} - g_{\mu\nu} \Box\right]f'(R)= T_{\mu \nu },
\label{field_eq}
\end{equation}
where $\displaystyle{T_{\mu\nu}= \frac{-2}{\sqrt{-g}}\frac{\delta\left(\sqrt{-g}{\cal L}_m\right)}{\delta g^{\mu\nu}}}$ is the energy momentum tensor of matter and the prime indicates derivative with respect to $R$.
The adopted signature is  $\left(+,-,-,-\right)$.
The dynamical system is completed by considering also  the contracted Bianchi identities
\begin{equation}
\nabla^{\mu}T_{\mu\nu}=0\,.
\label{bianchi}
\end{equation}
Enlarging the geometric sector   can be useful in view of  the dark side  since the further gravitational degrees of freedom have, in principle,   a role in addressing dark energy and dark matter issues \cite{uno3,annalen}.

Varying the action with respect to the metric is not the only choice. It is also possible to vary  with respect to the
affine connection, considering it independent from the metric itself \cite{nuovoo, ad551}. This is the so called \emph{Palatini approach} that produces different field equations. The advantage of the latter approach is that field equations remain of  second order  \cite{ad55, ad551,bhp}.
According to the Palatini formalism, an important remark is in order. The Ricci scalar
 is $R\equiv R( g,\Gamma)
=g^{\alpha\beta}R_{\alpha \beta}(\Gamma )$ being a
\textit{generalized Ricci scalar} and $ R_{\mu \nu }(\Gamma )$ is
the Ricci tensor of a torsion-less connection $\Gamma$, which,
{\it a priori}, has no relations with the metric $g$ of spacetime.
 Field equations, derived  from the Palatini variational principle are:
\begin{equation}
f^{\prime }(R)R_{(\mu\nu)}(\Gamma)-\frac{1}{2}f(R)g_{\mu \nu }=
T_{\mu\nu}\label{ffv1}
\end{equation}
\begin{equation}
\nabla _{\alpha }^{\Gamma }(\sqrt{-g}f^{\prime}(R)g^{\mu \nu })=0
\label{ffv2}
\end{equation}
where  $\nabla^{\Gamma}$ is the covariant derivative with respect
to $\Gamma$.   We  denote $R_{(\mu\nu)}$ as the symmetric part of
$R_{\mu\nu}$, i.e. $R_{(\mu\nu)}\equiv
\frac{1}{2}(R_{\mu\nu}+R_{\nu\mu})$.

In order to get (\ref{ffv2}), one has to additionally assume that the above
${\cal L}_m$ is functionally independent of $\Gamma$; however it may
contain metric  covariant derivatives $\stackrel{g}{\nabla}$ of the
fields. This means that the matter stress-energy tensor
$T_{\mu\nu}=T_{\mu\nu}(g,\Psi)$ depends on the metric $g$ and
matter fields  $\Psi$, and  their
derivatives  with respect to the Levi-Civita
connection of $g$.  It is  natural  to define a new
metric $h_{\mu \nu}$, such that
\begin{equation}\label{h_met}
\sqrt{-g}f^{\prime }(R)g^{\mu \nu}=\sqrt{-h}h^{\mu \nu }\,.
\end{equation}
This choice impose $\Gamma$
to be the Levi-Civita connection of $h$ and the only restriction
is that $\sqrt{-g}f^{\prime }(R)g^{\mu \nu}$ is
non-degenerate. In the case of Hilbert-Einstein Lagrangian, it is
$f^{\prime}(R)=1$ and the statement is trivial.

In both metric and Palatini  approaches, the function $f(R)$ is not fixed \emph{a priori}. Thus, its determination, according to data and phenomenology,   represents the main challenge for the $f(R)$ theory \cite{prico}.

From a different point of view, reliable classes of $f(R)$  models should be constrained at fundamental level \cite{lambiase}. Specifically, bounds on $f(R)$ models could be derived  by taking into account different vacuum states via Bogolubov transformations \cite{quantum,quantum1,quantum2}.
In fact,  in quantum field theories, the  Bogolubov coefficients drive  the different choices of vacuum states.
So, requiring that different classes of $f(R)$ functions change vacuum states according to Bogolubov transformations is a basic requirement to guarantee the $f(R)$ viability at fundamental level. This procedure  somehow fixes   the $f(R)$ free parameters and so it is of some help in  reconstructing  the  $f(R)$ form by means of basic requirements of quantum field theory \cite{ubi1,ubi2}. To this end, one has to confront  with the problem of  quantizing the space-time in a curved background and then to provide relations  between  quantization and $f(R)$ gravity at least at semiclassical  level. Hence, the Bogolubov coefficients allow to pass from a vacuum state to another through a semiclassical procedure  where the rate of particle production is minimized. If the rate is minimized,  one can fix, indeed,  the free parameters of a given  $f(R)$ model.  We assume that the rate is minimized to be consistent with cosmological high redshift observations, from one side, and with quantum field theory predictions, from the other side.

Furthermore, one can  relate the rate of particle production with the Hubble parameter and thus with the redshift $z$. In so doing,  it is possible to frame the Bogolubov coefficients in terms of observable cosmological quantities as $H_0$, the today observed Hubble constant, or  $R_0\sim \rho_0$,  the value of the today curvature or density.

The paper is organized as follows. In Sec. \ref{secy}, we sketch  the derivation of  the  Bogolubov coefficients as  semiclassical quantities  in the context of quantum field theory. In Sec. \ref{sez2} the rate of particle production is investigated  assuming a constant (de Sitter) curvature $R_0$ in the framework of non-minimally coupled theories of gravity. Such theories are the prototype of $f(R)$ models and then a generalization is straightforward. In Sec. \ref{sez3}, we minimize the  Bogolubov coefficients to   get constraints for  $f(R)$ free parameters in both metric and Palatini formalism. Applications to some  cosmological models are discussed. Outlooks and conclusions are reported in Sec. \ref{conclusions}.


\section{A semiclassical approach for particle production rate}\label{secy}

A strategy to derive  the  particle production rate in curved space  is to fix a background with a constant curvature i.e. $R=R_0$. This situation is usually named  as the {\it de-Sitter phase} \cite{birrell, ad1}. From the above field Eqs. (\ref{field_eq}), it is easy to derive  an  effective cosmological constant term  ${\displaystyle \Lambda_{eff}=\frac{f(R_0)}{2 f'(R_0)}=\frac{R_0}{4}}$, which, in principle, depending on the value of $R_0$, can give rise to an  accelerating expansion phase \cite{darkall}. The choice of  $R_0$   allows to simplify the calculations thanks to the symmetries of de Sitter spacetime.
In order to constrain the  form of $f(R)$ function, a  method is to fix the range of  free parameters by the  transition to  different vacuum states. Such a procedure relies on the definition of  the Bogolubov coefficients. To define them,  let us consider the quantization on  a curved background.

Since we are considering  modified theories of gravity,  a model where a scalar field $\phi$ is non-minimally coupled to geometry, i.e.  $\propto R\phi$ can be assumed. The related Klein-Gordon equation is
\begin{equation}\label{kbj}
\left[\Box-m^2+\xi R(x)\right]\phi=0\,,
\end{equation}
where $m$ is the effective mass of the field, $\xi$ is the coupling\footnote{It is worth noticing that any $f(R)$ theory of gravity can be recast as a non-minimally coupled theory through the identification $\phi\rightarrow f'(R)$ and the coupling  $f'(R)^{-1}$.}.

The general solution can  be expressed as a complete set of mode-solutions for the field $\phi$ \cite{iser}:
\begin{equation}\label{ret}
\phi(x)=\sum_i[a_i u_i(x)+a_{i}^{\dag}u_{i}^{*}(x)]\,,
\end{equation}
where it is possible to  adopt a specific set of mode solutions $u_{i}(x)$, albeit it is always possible to rewrite $\phi(x)$ for a different set
$\bar{u}_{j}$ as
\begin{equation}\label{diffset}
\phi(x)=\sum_j[\bar a_j \bar u_j(x)+\bar a_{j}^{\dag}\bar u_{i}^{*}(x)]\,.
\end{equation}
In other words, one can pass through different decompositions of $\phi$, defining a corresponding form of the vacuum solution that is, in general,   $\bar a_{j}|0\rangle\neq 0$,
in a curved space background. In fact, expressing the new modes, $\bar u_j$ in
terms of the old ones $u_i$, one can write
\begin{eqnarray}
\bar u_j&=&\sum_i (\alpha_{ji}u_i+\beta_{ji}u_{i}^{*})\,, \nonumber\\
u_i &=& \sum_j (\alpha_{ji}^{*}\bar u_j-\beta_{ji}\bar u_{j}^{*})\,,
\end{eqnarray}
where $\alpha_{ji}$ and $\beta_{ji}$ are defined as $\alpha_{ij}=(\bar
u_{i},u_j)$, $\beta_{ij}=-(\bar u_{i},u_{j}^{*})$ and satisfy the relations
\begin{eqnarray}\label{pro}
\sum_k
(\alpha_{ik}\alpha_{jk}^{*}-\beta_{ik}\beta_{jk}^{*})&=&\delta_{ij}\,,\nonumber\\
\sum_k (\alpha_{ik}\beta_{jk}-\beta_{ik}\alpha_{jk})&=&0\,.
\end{eqnarray}
In particular, if we consider the vacuum $|0\rangle$ then $a_{i}|0\rangle=0$, $\forall
i$, but, in general, it is  $a_{i}|\bar 0\rangle\neq 0$. Thence, it results from the definition of the particle number  given by $N_{i}\equiv
a_{i}^{\dag}a_i$, that is
\begin{equation}\label{jdnf}
\langle \bar 0 | N_{i} | \bar 0 \rangle = \sum_{j}|\beta_{ji}|^{2}.
\end{equation}
It follows that the physical meaning of such coefficients is associated to the rate of particle production. In fact,  generic coefficients $\beta_{ji}$ are associated to the particle number count  for given set of modes $i$. Specifically, $\alpha_{ji}$ and $\beta_{ji}$ are  referred to as the {\it Bogolubov coefficients} which identify the {\it Bogolubov transformations} and allow to pass from a vacuum state to another one.

Since the form of the $f(R)$ function is not known \emph{a priori}, by adopting the above semiclassical procedure and evaluating the different vacuum states for some classes of $f(R)$, we can minimize the rate of particle production. In so doing, we can constrain  the free parameters of a given $f(R)$ model. In particular, we will see that Bogolubov coefficients strictly depend on the form of $f(R)$.


\section{Particle production in non-minimally coupled theories of gravity}
\label{sez2}

The generic prototype of  alternative theories  of gravity are the non-minimally coupled scalar-tensor theories. As we said before, $f(R)$ theories and any extended theory can be recast as General Relativity with some  non-minimal coupling and  a further contribution in the stress-energy momentum tensor (see \cite{lobo} for the general procedure).
In this section, we discuss the Bogolubov transformations in the context of  homogeneous and isotropic cosmologies, in order to define the rate of particle production and then to constrain  the functional form of  $f(R)$ gravity.

\noindent The  particle production rate is a mostly universal feature, in the sense that it has not to depend on the particular gravitational background. Indeed, assuming a different  gravitational theory\footnote{For example, $f(R,G)$, General Relativity, $f(T)$, and so forth.} we expect that it is the same and can be consistently used to fix the parameters of the theory itself. In other words, this property is extremely relevant since it allows to consider the Bogolubov transformations  for different gravitational backgrounds. For the purposes of this work, we limit to the case of $f(R)$ gravity.

As we said, the simplest choice to construct the Bogolubov transformations is assuming  a de Sitter phase  with a constant curvature $R_0$. A spatially flat Friedman-Robertson-Walker (FRW) conformal metric is \cite{repo}
\begin{equation}\label{FRW_Conformal}
ds^2=\frac{1}{H^2\eta^2}\left(d\eta^2-dx_1^2-dx_2^2-dx_3^2\right),
\end{equation}
where we adopted the conformal time ${\displaystyle \eta=-\frac{1}{Ha(t)}}$, which varies in the interval $-\infty<\eta<0$. Introducing a scalar field $\phi(\eta ,\textbf{x})$ depending  on $\eta$ and $\textbf{x}\equiv(x_1,x_2,x_3)$, the corresponding Klein-Gordon equation Eq. (\ref{kbj}), in terms of   space-time modes, can  be rewritten as
\begin{equation}\label{lhjefilejlo}
(\Box-m^2+\xi R)\phi(\eta ,\textbf{x})=0.
\end{equation}
This equation is formally equivalent to Eq. (\ref{kbj}), albeit the  functional dependence on the  variables $\eta$ and $\textbf{x}$ is explicit. Without choosing $\xi$ \emph{a priori},  the corresponding class of solutions is
\begin{equation}\label{kdjfhkfhkhfkjdwhfdkjh}
\phi(\eta,\mathbf{x})=\phi_k(\eta) e^{i \left(\mathbf{k}\cdot
\mathbf{x}\right)},
\end{equation}
where the wave vector is decomposed as $\mathbf{k}\equiv (k_1,k_2,k_2)$.
By scaling ${\displaystyle \phi(\eta ,\textbf{x})=\frac{\tilde{\phi}(\eta ,\textbf{x})}{a}}$,  the Klein-Gordon differential equation for the FRW metric \eqref{FRW_Conformal} is
\begin{equation}\label{equazionenuova}
\tilde{\phi}_k''(\eta)+\omega^2(\eta,k,\xi)\tilde{\phi}_k(\eta)=0\,,
\end{equation}
where the prime stands for the derivative with respect to the conformal time  $\eta$.

\begin{center}
\begin{figure}
\flushleft
\includegraphics[scale=0.6]{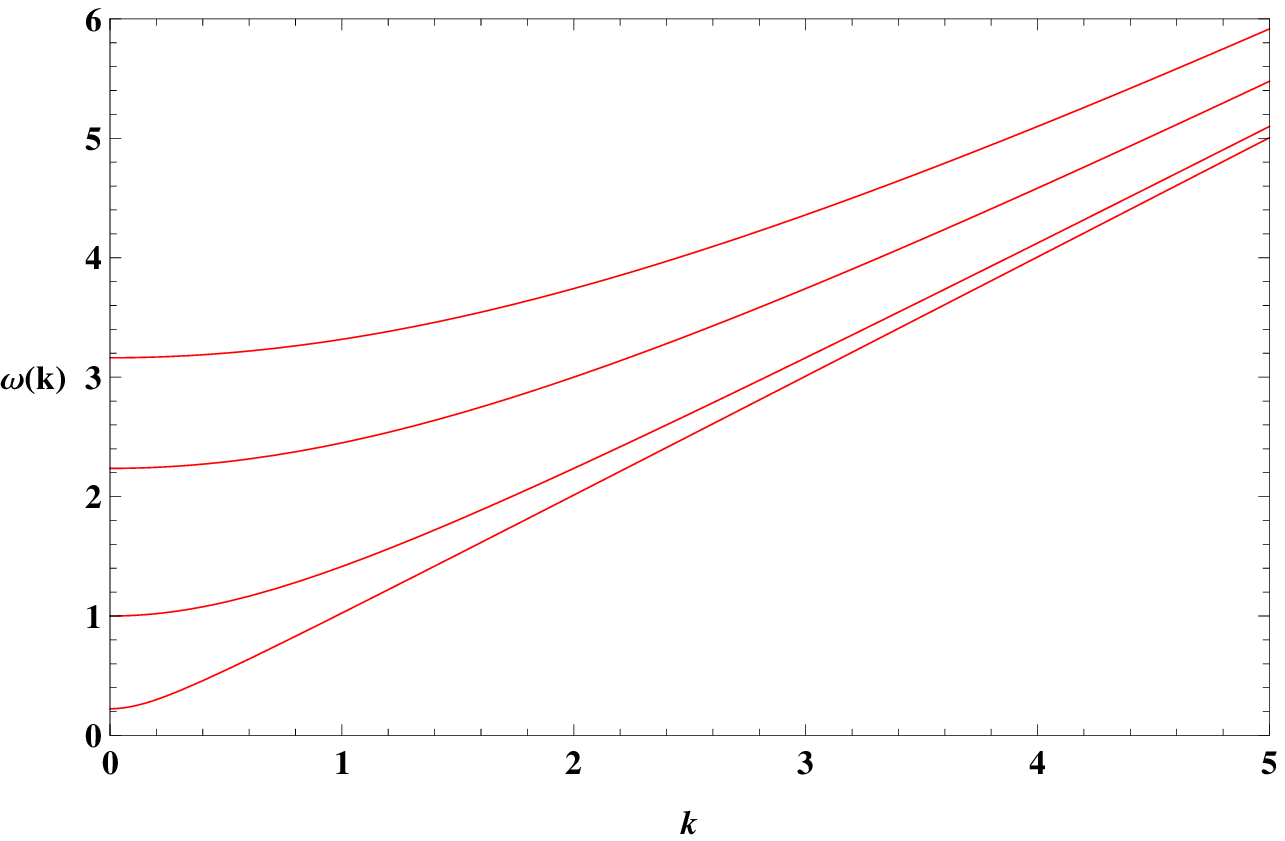}
\includegraphics[scale=0.6]{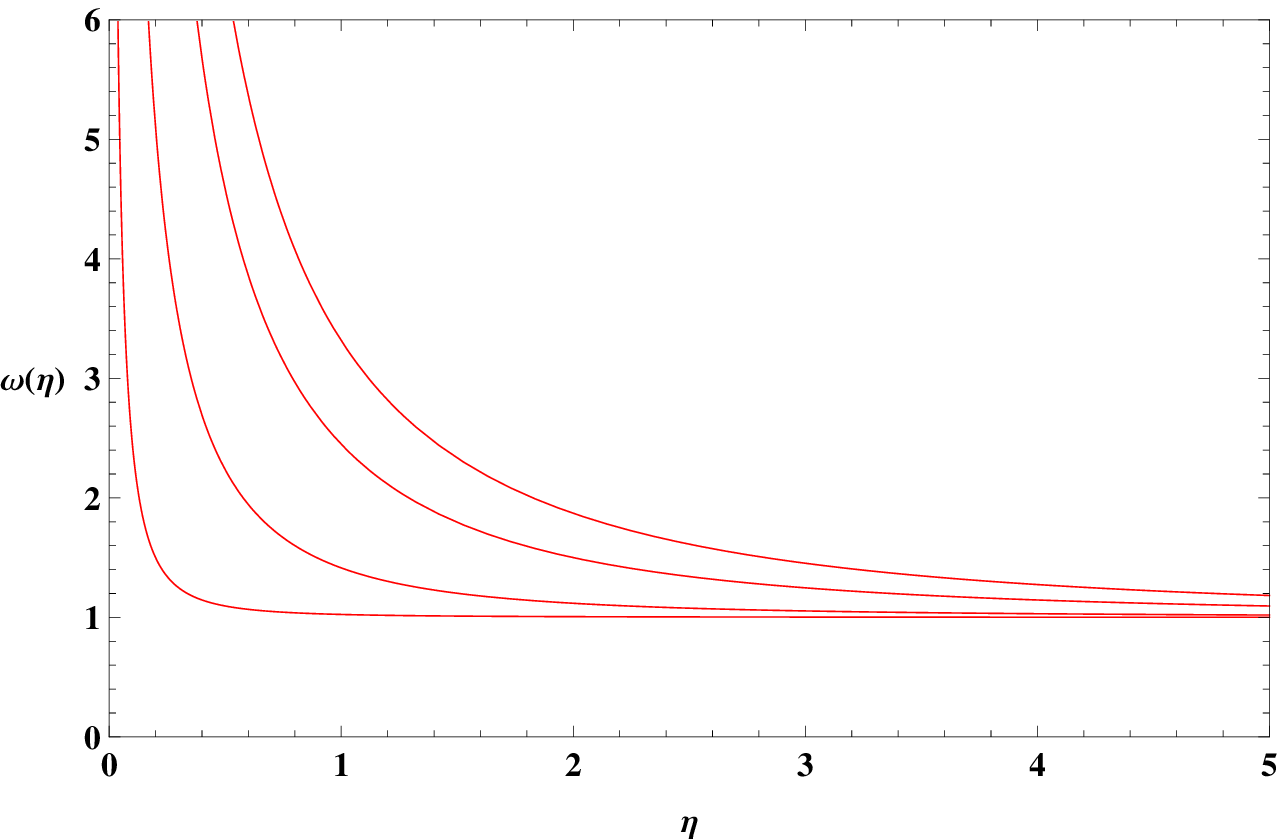}
{\caption{Plots of the $\omega$ profiles in function of $k$ and $\eta$ respectively on the left and right. The left plots show the $\omega(k)$ behaviors for four indicative ratios $\frac{M_{eff}^{2}}{H^2\eta^2}=0.05,1,5,10$. The figures on the right, on contrary, show the $\omega(\eta)$ behaviors in function of $\eta$, with indicative ratios $\frac{M_{eff}^{2}}{H^2}=0.05,1,5,10$ and conventionally with $k=1$.
}
\label{omega1e2}}
\end{figure}
\end{center}

\noindent The above equation is analogue  to the harmonic oscillator with $\omega$ depending on the conformal time $\eta$. The $\omega$ parameter  takes the form:
\begin{equation}
\omega(\eta)=\sqrt{k^2+ a^2\Big[m^2+ 2f(\xi) H^2\Big]}\,,
\end{equation}
where $f(\xi)\equiv6\xi -1$. For our purposes,  the function $f(\xi)$ can be conventionally positive-definite assuming $\xi>{1\over6}$. Moreover, it is convenient to define an  effective mass $M_{eff}$ as
\begin{equation}\label{jkhknmbmnb}
\frac{M_{eff}^2}{H^2}\equiv \frac{m^2}{H^2}+2f(\xi)\,,
\end{equation}
where $m$ is the state of mass of the scalar field  and  ${\displaystyle H\equiv{\dot a \over a}}$ is the Hubble parameter. Since ${\displaystyle \xi>{1\over6}}$,  $M_{eff}$ is always positive. The frequency dependence is
\begin{equation}\label{dwkljlewjrlj}
\omega(\eta)=\sqrt{k^2+\frac{M_{eff}^{2}}{H^2\eta^2}}.
\end{equation}
which is always positive for $\xi\geq{1\over 6}$. The functions  $k(\eta)$ and $\omega(\eta)$ are sketched in Fig. (\ref{omega1e2}) for some cases of interest.

\noindent The general solution  $\tilde{\phi}_k(\eta)$ reads
\begin{equation}\label{hghjghj}
\tilde{\phi}_k(\eta)=\sqrt{\eta}\left(A_{k}H_{k,\nu}^{(1)}(\eta)+B_{k}H_{k,\nu}^{(2)}(\eta)\right),
\end{equation}
where $H_{k,\nu}^{(1)}$ and $H_{k,\nu}^{(2)}$ are Hankel's functions of first and second type respectively, with the  position \cite{prunz,prunz1}
\begin{equation}\label{msa}
\nu\equiv\sqrt{\frac{1}{4}-\frac{M_{eff}^{2}}{H^{2}}}.
\end{equation}
The corresponding asymptotic behavior is relevant to  infer the particle production rate.
In the case $\eta\rightarrow 0^{-}$, we obtain
\begin{equation}\label{lj}
\tilde{\phi}_k(\eta)\rightarrow
\frac{\sqrt{|\eta|}}{\pi \nu}\left\{\sin(\pi\nu)\Gamma(1-\nu)\left(\frac{k\eta}{2}\right)^{\nu}-\Gamma(1+\nu)\left(\frac{k\eta}{2}\right)^{-\nu}\right\},
\end{equation}
and the square modulus of $\beta_k$ is \cite{pranz,pranz1}
\begin{equation}\label{jkkjhkkhkj}
|\beta_k(\eta)|^2=\frac{\omega_{k}}{2}|\phi_k(\eta)-\frac{i}{\omega_{k}(\eta)}\dot\phi_k(\eta)|^{2}.
\end{equation}
We are interested in the case ${\displaystyle \frac{M_{eff}}{H}\gg 1}$, which corresponds either to the situation where the effective mass dominates over the Hubble rate or $H$ is small at late times of the universe evolution. Thus, we find out
\begin{equation}\label{,lfj}
|\beta_k(\eta)|^2 \sim \frac{H^{3}}{32\pi
m^3}|\Gamma\left(1-i\frac{m}{ H}\right)|^2\exp(\pi m),
\end{equation}
where  $\Gamma$ is  the Euler function.

\begin{center}
\begin{figure}[h!]
\flushleft
\includegraphics[width=%
0.5\columnwidth]{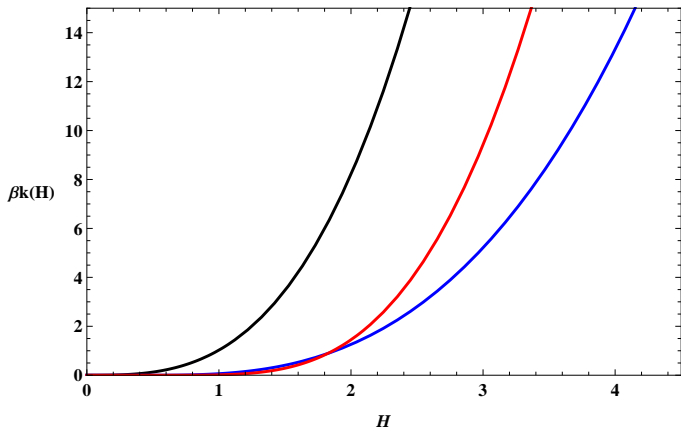}
{\caption{Plot of the Bogolubov coefficient $|\beta_k|^2$ varying in the range $H=0\ldots5$, with $m=0.01;1;2$ respectively for the black, blue and red lines.
}
\label{bogo}}
\end{figure}
\end{center}

\noindent In the case ${\displaystyle \frac{M_{eff}}{H}\ll 1}$, we obtain  that the Bogolubov coefficients are negligibly small \cite{pronz}, that is
\begin{equation}\label{kfhdkfh}
|\beta_k(\eta)|^2\ll 1.
\end{equation}
Hence, by assuming that ${\displaystyle \frac{M_{eff}}{H}\ll 1}$, we can consider two different cases. The first is $H\gg m$, with $m\rightarrow 0$. The second is  un-physical, since it provides a diverging Bogolubov coefficient $\beta_k$ \cite{kolb}. Thus, by assuming the validity of the above results, we are able to relate the $f(R)$ gravity to Bogolubov coefficients constraining   the free parameters of the models.  To this goal, we assume to pass through different vacuum states. Clearly, different $f(R)$ gravity models means different couplings $\xi$.


\section{Minimizing the rate of particle production in $f(R)$ gravity}
\label{sez3}

Let us focus on the physical case ${\displaystyle \frac{m}{H}\ll1}$ in the context of $f(R)$ gravity. We investigate such a case in the  de Sitter phase with $R=R_0$ and  minimize the Bogolobuv coefficients obtaining,  correspondingly, the minimum of   particle production rate. 

\noindent  Such a quantity has to  be minimized essentially for two  reasons. The first concerns cosmological observations at high energy regimes. For example, taking into account the cosmic microwave background,  cosmological measurements could not be compatible with particle   production rate, so  the condition ${\displaystyle \frac{m}{H}\ll1}$ is required to guarantee that any theory of  gravity works at  high $z$. In addition,  assuming to pass from different vacuum states, it is important to test if cosmological models describe the vacuum according to observations. In this perspective, minimizing Bogolubov coefficients is a powerful tool to discriminate among different models (see also \cite{planck}).

In particular, our purpose is to infer physical bounds on the free parameters of some classes of $f(R)$ models  by Bogolubov transformations. The first step is to write $H$ from the cosmological equations derived in the $f(R)$ framework. In a FRW universe we obtain in  metric
and Palatini formalism respectively:
\begin{subequations}\label{dojfdj}
\begin{align}
H^2 &= \frac{1}{3} \left[ \rho_{curv} + \frac{\rho_m}{f'(R)}\right],
\\
H^2 &=
\frac{1}{6f'(R)}\left[\frac{2 \rho+Rf'(R)-f(R)}{G(R)}\right].
\end{align}
\end{subequations}
where $\rho_m$ is the standard matter  density.  The effective curvature density term is \cite{anc2}
\begin{equation}\label{argh}
\rho_{curv} = \frac{1}{2}\left[\frac{f(R)}{f'(R)}-R\right]-3H\dot{R}\left[ \frac{f''(R)}{f'(R)}\right]\,,
\end{equation}
and  the function $G(R)$ is given by
\begin{equation}\label{lkjdf}
G(R)=\left[ 1-\frac{3}{2}\frac{f''(R)(Rf'(R)-2f(R))}{f'(R)(Rf''(R)-f'(R))}\right]^2\,.
\end{equation}
As said before, clearly $f(R)$ gravity can be recast in term of a scalar-tensor theory as soon as the identifications $\phi\rightarrow f'(R)$, for the field,  and $G_{eff}\rightarrow f'(R)^{-1}$, for the coupling,  are adopted (see \cite{libroSV} for an extended discussion on this point).

\begin{center}
\begin{figure}[h!]
\flushleft
\includegraphics[scale=0.42]{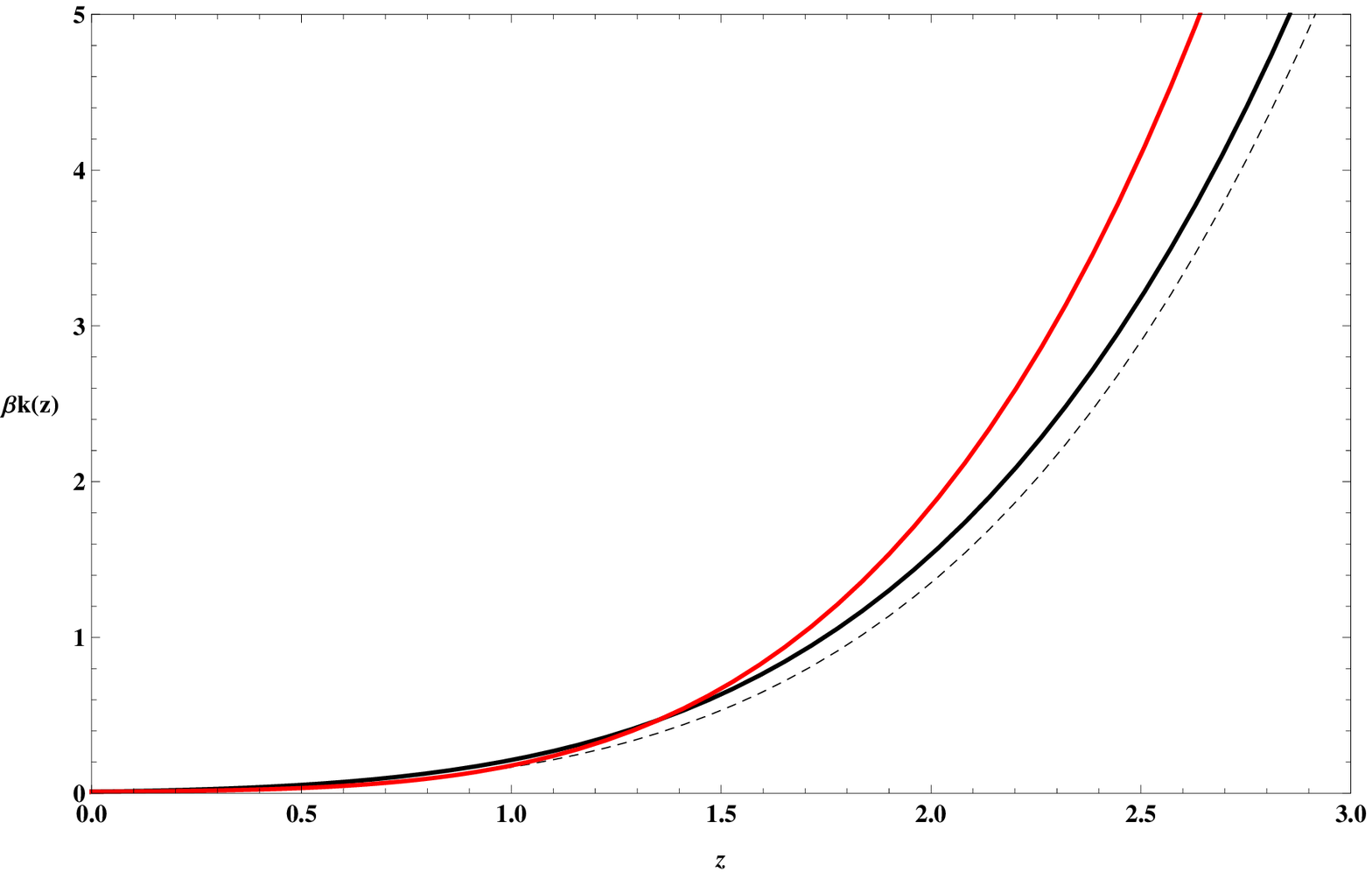}
\includegraphics[scale=0.42]{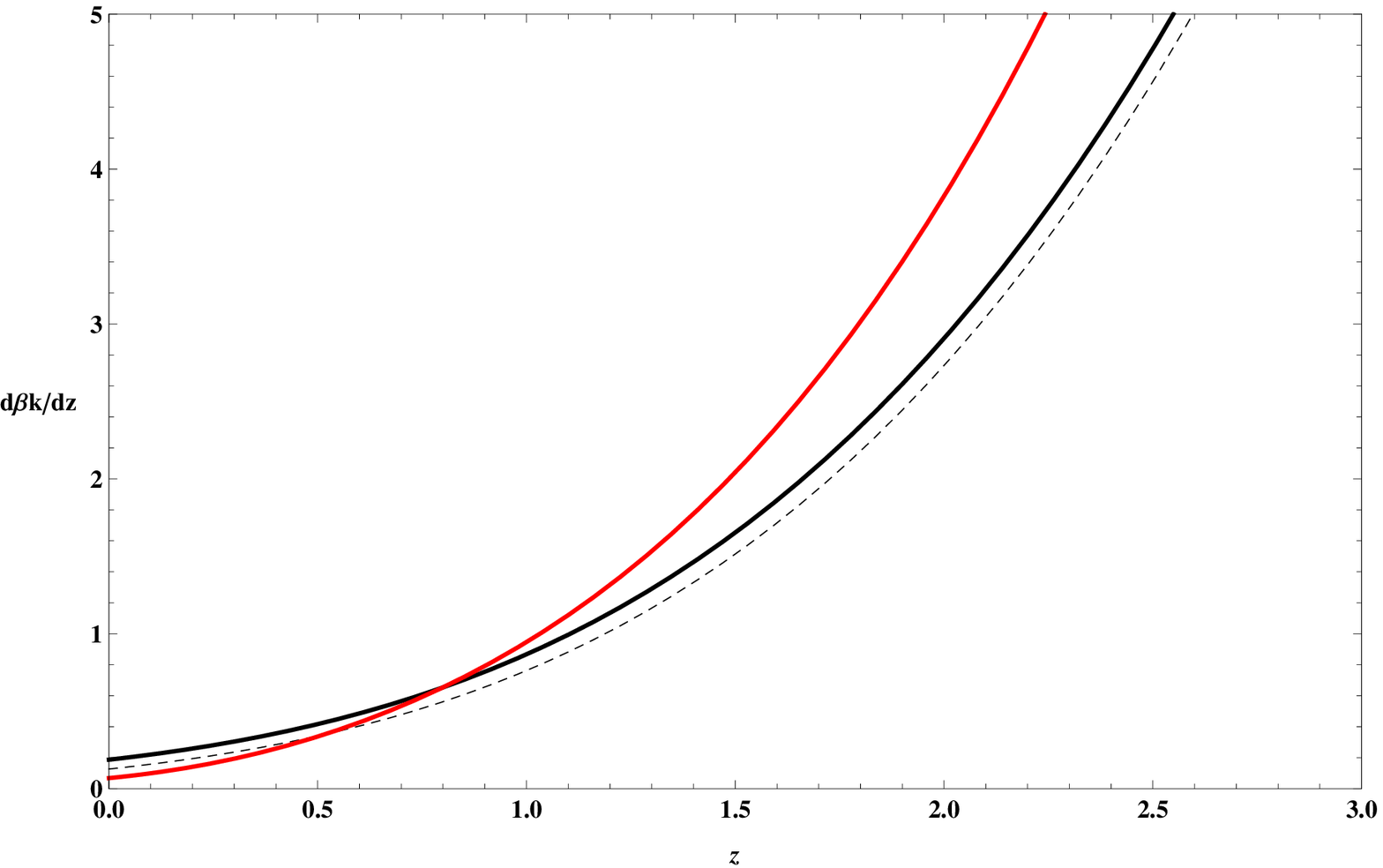}
{\caption{Bogolubov coefficients (left figure) evaluated for the $\Lambda$CDM model (black line), the cosmographic expansion (dashed line) and the Chaplygin gas (red line), with matter density $\rho=0.27$, and the Chaplyging gas coefficients: $A=0.9$, $\beta=0.8$, (see \cite{chaply,chaply1,chaply2}) with the normalized Hubble rate $H_0=0.68$. Derivatives of $\beta_k(z)$ (right figure) have been reported for the same cases, i.e. $\Lambda$CDM, cosmographic and Chaplygin gas.
}
\label{bogo}}
\end{figure}
\end{center}
Besides, the particle  production rate can  be achieved at first order by a Taylor expansion of the Bogolubov coefficient $\beta_k$. It is
\begin{eqnarray}\label{mater}
|\beta_{k}|^2=e^{\pi m}\Big[\frac{H^{3}}{32\pi
m^3}+\gamma ^2\frac{H}{32\pi m}\Big],
\end{eqnarray}
where we adopted the $\Gamma(1-ix)$ function and its Taylor series in case $x\ll1$, obtaining $\Gamma\sim 1+i\gamma x$. The constant $\gamma$ is  the Euler constant and reads $\gamma\sim0.577$.

As a first step, one can compare such  Bogolubov coefficients with the Hubble rate expressed as function of the redshift $z$. In such a way, the form of $\beta_k$ becomes a function of the redshift as well. This has been reported in the left plot of Fig. III, whereas in the right plot we draw the variation of $\beta_k$ as the redshift increases, i.e. its first derivative with respect to the redshift $z$. The reported three models are: $(i)$ the  $\Lambda$CDM model  \cite{lambda}; $(ii)$ a cosmographic expansion where the  deceleration parameter variation, namely the jerk parameter,  is  $j(z)\geq1$  \cite{ratra};  $(iii)$ the Chaplygin gas where dark energy and dark matter are considered under the standard of  a single fluid \cite{chaply,chaply1,chaply2}. These models can be considered as   the most relevant paradigms for describing dark energy \cite{yoo}.

In the case of constant curvature $R=R_0$  related to  a de-Sitter phase, we obtain
\begin{eqnarray}\label{k,n,nx,mcn,n,n}
|\beta_{k}|^2&=&e^{\pi m}\left[\frac{1}{3^{\frac{3}{2}}32\pi m^3}\left(\frac{\rho_0}{f_{0}^{'}}-\Lambda_{eff}\right)^{3/2}+\frac{\gamma^2}{32 \sqrt{3} \pi m }\left(\frac{\rho_0}{f_{0}^{'}}-\Lambda_{eff}\right)^{1/2}\right]\,,
\end{eqnarray}
and
\begin{eqnarray}
|\beta_{k}|^2&=&e^{\pi m}\left[\frac{1}{32\pi
m^3}\left(\frac{(2\rho_0+R_0-2\Lambda_{eff})(R_0-\frac{f^{'}_{0}}{f_{0}^{''}})}{R_0-\frac{f^{'}_{0}}{f_{0}^{''}}-\frac{3}{2}R_0+6\Lambda_{eff}}\right)^{3/2}+\frac{\gamma ^2}{32\pi
m}\left(\frac{(2 \rho_0+R_0-2\Lambda_{eff})(R_0-\frac{f^{'}_{0}}{f_{0}^{''}})}{R_0-\frac{f^{'}_{0}}{f_{0}^{''}}-\frac{3}{2}R_0+6\Lambda_{eff}}\right)^{1/2}\right]\,
\end{eqnarray}
respectively for metric and Palatini formalism. Hereafter $f_0\equiv f(R=R_0)$,  $f'_0\equiv f'(R=R_0)$, and $f''_0\equiv f''(R=R_0)$ and $\rho_0$ is the value of standard matter-energy density for $R_0$.  Since the form of $f(R)$ is not known \emph{a priori}, we need to consider  cases of particular interest \cite{anc3,odint2003} as
\begin{subequations}\label{trans}
\begin{align}
f_1(R)&= R^{1+\delta}+\Lambda\,,\\
f_2(R)&= R+\epsilon R^2\ldots\,,\\
f_3(R)&= R+ R^n+\sigma R^{-m}\,,\\
f_4(R)&= R-\frac{\alpha(R)^n}{1+\beta(R)^n}\,.
\end{align}
\end{subequations}
We therefore need to fix via Eq.  The coefficients $\delta,\epsilon,\sigma,n,m,\alpha$ and $\beta$ can be fixed by Eq.(\ref{mater}). To do so, we require that the rate of particle production is negligible or essentially as small as possible \cite{anc4}. Hence, the strategy to follow is to assume that the free parameters of Eqs. ($\ref{trans}$) minimize the Bogolubov coefficients. Following this procedure, we obtain the results of Tabs. I and II.

\begin{table}[h]
\begin{center}
\begin{tabular}{|c|c|c|}
\hline\label{snstc}
$f_n(R)$ & $num. param. $ & $minimiz.$\\
\hline\hline\hline
$f_{1M}(R)$ &  $1$ & $1+\delta\leq\frac{4\rho_0}{R_{0}^{1+\delta}}$\\
\hline
$f_{2M}(R)$ & $1$ & $\epsilon\leq\frac{2\rho_0}{R_{0}^{2}}-\frac{1}{2R_{0}}$ \\
\hline
$f_{3M}(R)$ &  $3$ & $nR_{0}^{n}\Big[1-{m\over n}\sigma R_0^{-(m+n)}\Big]\leq4\rho_0-R_0$\\
\hline
$f_{4M}(R)$ &  $3$ & $\left(R_{0}-n R_{0}^n\alpha+2R_{0}^{1+n}\beta+R_{0}^{1+2n}\beta^2\right)\left(1+R_{0}^n\beta^2\right)^{-1}\leq 4\rho_0-R_0$\\
\hline\hline
\end{tabular}
\caption{Table of minimizing conditions for the free parameters of $f(R)$ models from Eq. ($\ref{trans}$) in the metric formalism. Here the subscript \emph{M} stands for \emph{metric}. The above equalities correspond to the case of vanishing Bogolubov coefficients, whereas the inequalities to more general cases where the Bogolubov coefficients are not  zero.}
\end{center}
\end{table}

\begin{table}[h]
\begin{center}
\begin{tabular}{|c|c|c|}
\hline\label{snstc2}
$f_n(R)$ & $num. param. $ & $minimiz.$\\
\hline\hline\hline
$f_{1P}(R)$ &  $1$ & $\forall\delta<0,\,\,\bigvee\,\,\delta\geq1,\,\, \delta\neq-1$\\
\hline
$f_{2P}(R)$ & $1$ & $\epsilon<0$ \\
\hline
$f_{3P}(R)$ &  $3$ & $\left(1+nR_{0}^{n-1}-m\sigma R_{0}^{-(m+1)}\right)\left(n(n-1)R_{0}^{n-2}+m(m+1)\sigma R_{0}^{-(m+2)}\right)^{-1}\leq R_0$\\
\hline
$f_{4P}(R)$ &  $3$ & $R_{0}^{1-n}(1+R_{0}^n \beta)\left(R_{0}+R_{0}^{1+2n}\beta^2+R_{0}^n(2\beta R_{0}-\alpha n)\right)\left(n\alpha(1-n+(1+n)R_{0}^n\beta)\right)^{-1}\leq R_0$\\
\hline\hline
\end{tabular}
\caption{Table of minimizing conditions for the free parameters of $f(R)$ models from Eq. ($\ref{trans}$) in the Palatini formalism. Here the subscript \emph{P} stands for \emph{Palatini}. Inequalities and equalities follow the same considerations of Tab. I. Here, we assumed that $\rho_0+{R_0\over 4}>0$.}
\end{center}
\end{table}
One  can  calibrate the constraints  over the free parameters  in Tabs. I and II using also late-time and CMBR cosmological constraints \cite{ratra,planck}.
In general, any consistent  choice of $f(R)$ gravity leads to
\begin{equation}\label{lr}
f^{'}(R)\leq\frac{4\rho_0}{R_0}\,,
\end{equation}
in the metric formalism
(where the equality requires  vanishing Bogolubov coefficients) and
\begin{subequations}
\begin{align}
R_0&\geq-4\rho_0\,,\label{primadi}\\
f^{'}_{0}&\neq R_0f^{''}_{0}\label{secondadi}\,,
\end{align}
\end{subequations}
in the Palatini formalism. Again, the equalities lead to  vanishing  Bogolubov coefficients. In addition, Eq. \eqref{primadi} represents a natural constraint on $R_0$.
These  conditions have to  be satisfied,  if one wants to pass through different vacuum states, without a significant particle  production rate. In principle, once evaluated the above constraints, it would be also possible to numerically constrain the derivatives of $f(R)$ models. For example, to guarantee that the gravitational constant does not significantly  depart  from the Solar System limits, one needs that $4\rho_0\sim R_0$. Thus, observations on $\rho_0$ open the possibility to constrain $R_0$ and may be compared to cosmological constraints over $R_0$ itself \cite{cosmogra,cosmogra1}. Analogously, in the Palatini case, $R_0$ is somehow comparable to $-4\rho_0$. Hence, a correct determination of the limits over $R_0$ could also discriminate between the metric and Palatini  approaches.

\section{Outlooks and perspectives}\label{conclusions}

We investigated the role played by the  particle production rate  in the context of $f(R)$ gravity. To this aim, we considered both the metric and Palatini approaches  reproducing   particle production  in both cases. Specifically, we derived  the Bogolubov coefficients, which permit to pass from a vacuum state to another. These coefficients can be related to the Hubble parameter  which strictly  depends on the  functional form  of a given  $f(R)$ model. In this sense, the particle production rate depends on the specific form of   $f(R)$ gravity. Hence, this is a method to constrain the free parameters of a given model, invoking a semiclassical scheme. Indeed, since the function $f(R)$ is not defined \emph{a priori}, it is necessary to determine some theoretical conditions on  $f(R)$  parameters at some fundamental level. Thus, we assumed to minimize the Bogolubov coefficients, i.e. the  particle production rate, allowing us to pass through different vacuum states, once postulated the background. in particular, we considered a  the de-Sitter phase $R=R_0$ and derived the Bogolubov transformations for some class of $f(R)$ models taking advantage from the fact that such theories can be easily recast as scalar-tensor models. The Bogolubov coefficients have been evaluated for a homogeneous and isotropic universe, postulating that the  particle production rate is negligibly small. This provided conditions on the form of $f(R)$ which have been reported in Tabs. I and II. As  result,  constraints can be derived  on  free parameters  of different classes of $f(R)$ functions. Such constraints can be combined  with  Solar System constraints  under suitable  conditions. In particular, we demonstrated that  cosmological measurements of $R_0$ would discriminate between  metric or Palatini approaches. A straightforward generalization of the  method would be to consider   the Bogolubov transformations for  space-times with variable curvature. Also in those cases one may check how to minimize the rate of particle production  in order to pass from a vacuum to another one. Furthermore, it would be possible to evaluate Bogolubov coefficients in other  modified gravity theories  to seek for constraints on free parameters \cite{frg}.  These topics will be the argument of future developments.

\section*{Acknowledgements}
SC and MP acknowledge financial support of INFN \emph{iniziative specifiche} QGSKY and TEONGRAV. SC thanks the TSPU for being awarded as  Honorary Professor.

\end{document}